\documentstyle[12pt,epsf,rotate]{article}

\def\baselinestretch{1.1}
\catcode`\@=11

\@addtoreset{equation}{section}

\def\underline#1{\relax\ifmmode\@@underline#1\else
        $\@@underline{\hbox{#1}}$\relax\fi}
\@twosidetrue

\long\def\@makecaption#1#2{
 \vskip 10pt \setbox\@tempboxa\hbox{#1: #2}
 \ifdim \wd\@tempboxa >\hsize #1: #2\par \else \hbox
        to\hsize{\box\@tempboxa\hfil} \fi}

\catcode`\@=12

\relax

\def\ltap{\raisebox{-.4ex}{\rlap{$\sim$}} \raisebox{.4ex}{$<$}}

\def\beq{\begin{equation}} \def\eeq{\end{equation}} 
\def\bea{\begin{eqnarray}} \def\eea{\end{eqnarray}} 
\def\bq{\begin{quote}} \def\eq{\end{quote}}  
\newcommand{\tabl}[2]{\mbox{$\displaystyle \frac{d#1}{d#2}$}}
\def\T{\textstyle}
\newcommand{\bvec}[3]{\left(\begin{array}{c} #1 \\ #2 \\ #3 \end{array}
\right)}
\newcommand{\tf}[2]{$\frac{\T #1}{\T #2}$}

\relax

\def\SMo{Standard Model}
\def\SM{\SMo\ }

\def\LRo{left--right} \def\LR{\LRo\ } 

\def\vr2{v_{R}^{2}}
\def\kap2{\kappa^{2}}
\def\kapp2{\kappa'^{2}}
\def\hvr{\hat{v}_{R}}
\def\hvr2{\hat{v}_{R}^{2}}
\def\hkap2{\hat{\kappa}^{2}}
\def\hkapp2{\hat{\kappa'}^{2}}

\begin{document}

\begin{titlepage}
  \renewcommand{\baselinestretch}{1}
  \renewcommand{\thefootnote}{\alph{footnote}}
  \thispagestyle{empty}

 \ \vskip -1.8cm

                     {\hfill TUM--HEP--241/96}

\vspace*{-.3cm}      {\hfill MPI--PhT/96--21}

  \vspace*{1.0cm}
  {\begin{center}       {\Large\bf
                        Gauge Coupling Unification in Left--Right\\ 
                        \ \\
                        Symmetric Models}
                        \end{center}  }
  \vspace*{1.cm}
  {\begin{center}       {\large
                        M. Lindner\footnote{\makebox[1.cm]{Email:}
                        Manfred.Lindner@Physik.TU-Muenchen.DE}
                        and
                        M. Weiser\footnote{\makebox[1.cm]{Email:}
                        Manfred.Weiser@Physik.TU-Muenchen.DE}}\\
                        \end{center} }
  \vspace*{-0.2cm}
  {\it \begin{center}  Institut f\"ur Theoretische Physik,
                       Technische Universit\"at M\"unchen,          \\
                    James--Franck--Strasse, D--85748 Garching, GERMANY
       \end{center} }
  \vspace*{3.cm}
{\Large \bf \begin{center} Abstract  \end{center}  }
We explore possibilities of gauge coupling unification in \LR 
symmetric models with non--minimal particle content. In addition to 
unification we require the absence of anomalies and sufficient proton
lifetime. Numerous previously unknown solutions are presented 
where unification occurs within the latest experimental errors. 
Solutions exist where the scale of \LR symmetry breaking can 
be as low as ${\cal O}(TeV)$ or the scale $M_{GUT}$ as high as the Planck
scale.

\ \vskip 3.cm
\hfill 
\begin{center}
{\small Work supported in part by EC grant ERB~SC1*CT000729
and DFG grant Li519/2--1}
\end{center}
\renewcommand{\baselinestretch}{1.2}

\end{titlepage}

\newpage
\renewcommand{\thefootnote}{\arabic{footnote}}
\setcounter{footnote}{0}

It is well known that gauge coupling unification is possible in the 
{\em minimal} \LR (LR) symmetric model \cite{LR1} if the scale of LR 
symmetry breaking, $M_R$, is chosen around $10^{10}$~GeV (see e.g., 
\cite{ShabanSt}). Attempts to lower or rise $M_R$ sizably require a 
different, non--minimal  particle content. Two possible directions 
for such changes which modify the $\beta$--functions (and thus gauge 
coupling unification) are: (i) supersymmetrization or 
(ii) new (``exotic'') fermionic and/or scalar representations.
In the supersymmetric case a few solutions for gauge coupling 
unification with low $M_R$ are already known. Deshpande, Keith and 
Rizzo \cite{DKR} investigated for example a SUSY--$SO(10)$ scenario 
(which is supersymmetric both in the LR and the \SM sector) and 
found at two--loop level unification for $M_R=1$~TeV at the scale
$M_X \approx 10^{16.2\pm0.4}$~GeV. They also studied
non--supersymmetric $SO(10)$--unification and concluded that one would
need a large number of Higgs doublets and triplets coming from the
scalar representations {\bf 16} and {\bf 126}, respectively. In
addition, $M_{GUT} = 10^{13.6}$~GeV would be much too low.
Two further solutions were recently presented by Ma, where the 
supersymmetric LR model is embedded into a larger supersymmetric 
gauge group: In one case \cite{Ma1} into the SUSY--$SO(10)$ 
(with additional discrete symmetries) and 
SUSY--$SO(10) \otimes SO(10)$, respectively, and in the other case 
\cite{Ma2} into the SUSY--$E6$. 
In both cases, however, new {\em exotic fermions} must be postulated  
in order to achieve gauge coupling unification for a low right--handed
scale.  We will discuss in this letter interesting solutions which 
emerge from new fermionic and scalar particles in \LR symmetric 
models even without supersymmetry. To our knowledge viable scenarios 
which achieve  unification even with $M_R={\cal O}(TeV)$ in 
{\em non}--supersymmetric \LR models were not published so
far.\footnote{We concentrate here on models where the scale of
LR--parity breaking is the same as $M_R$, i.e., the scale of
$SU(2)_R$--symmetry breaking. If one relaxes this condition solutions
with low $M_R$ yet exist (see, e.g., \cite{Mohapatra} where the
so--called D-parity is broken before $SU(2)_R$).}

Let us recall some attempts to modify the original minimal $SU(5)$ 
embedding of the \SM where the proton life time, $\tau_P$, and 
the weak mixing angle, $sin^2(\theta_W)$, are too small. 
Frampton and Glashow \cite{Frampton} proposed originally to add a few
chiral fermion multiplets in real representations with the same quantum
numbers as the known quarks and leptons. In this way it was possible to
increase $\tau_P$ and $\sin^2(\theta_W)$ by the necessary amounts. When
more precise data became available Amaldi, de Boer, Frampton,
F\"urstenau, and Liu \cite{Amaldi2} repeated the analysis of the gauge
coupling unification \cite{Amaldi1} for the \SMo, the minimal
supersymmetric \SM and an adequately modified $SU(5)$ with exotic
fermions and extra scalars. They investigated the running of gauge
couplings for different combinations of new multiplets at the
two--loop level where the contribution of exotic particles started at
a common effective mass scale. This threshold scale was treated as a
free parameter in the range of \hbox{$M_Z \leq M_{thres}\leq 10$~TeV}.
In this way they were able to find many unification solutions, even if
they added only four multiplets. In all of these cases there is no
contradiction with proton life time, and by using only real
representations potential chiral anomalies are automatically avoided.

Our aim is to perform a similar search for gauge coupling unification
in the class of \LR symmetric models. We will therefore extend the
minimal LR model and introduce new fermionic and scalar multiplets
which we call ``exotic''. The effects of these particles will be
switched on at a common effective threshold which is essentially
identical to the right handed scale $M_R$. We calculate the relevant
$\beta$--functions to one--loop order and restrict ourselves to cases
where this is sufficient. We search only for gauge unification
solutions with no specific GUT--scenario in mind. Therefore we pay no
attention to whether these representations fit into representations of
a specific GUT group. The requirements for a ``solution'' are
unification within the current $1\sigma$--error bars of inverse
couplings and sufficient proton life time. Furthermore we require the
absence of Landau poles before unification. Since we do not restrict
our search to real representations we must also pay attention to
potential anomalies.

This letter is organized as follows: First we introduce the notation
and simultaneously remind the reader about some features of the
minimal LR model. Next the anomaly coefficients and the
$\beta$--function contributions are determined for different new
``exotic'' particles and listed in tables. Finally we search for
solutions with the mentioned requirements and present numerous
previously unknown solutions for unification in non--minimal and
non--supersymmetric \LR models.

We study \LR symmetric extensions of the \SM with the gauge group 
$ G = SU(3)_c \otimes SU(2)_L \otimes SU(2)_R \otimes U(1)_{B-L}$.
Representations of this gauge group are written as
$(N_c, N_L, N_R, B-L)_{l,r}$, where $N_c$, $N_L$, $N_R$ are the 
dimensions under $SU(3)_c$, $SU(2)_L$ and $SU(2)_R$, respectively, 
where $B-L$ is the $U(1)_{B-L}$ quantum number, and where the index
``l'' or ``r'' specifies the chirality in case of fermionic
representations. Since we deal with non--minimal scenarios it is
important to remember that the indices ``L'' and ``R'' of the $SU(2)$
groups are not necessarily related to the chiralities ``l'' and ``r''
of the fermions. In the \SM and in the minimal \LR symmetric model the
chiralities ``l'' and ``r'' are uniquely related to one of the $SU(2)$
groups thus justifying to speak of ``left'' and ``right'' gauge groups.
In general, however, a direct connection between the chiralities of the
fermions and their quantum numbers is not required. We are thus free to
choose the chirality of any additional non--standard fermion
representation of the gauge group. In this way it is possible to
introduce, for example, bi--doublet fermions. Obviously ``$B-L$'' is
also no longer related to baryon and lepton number for such exotic
particles and we will therefore write ``$B-L+X$'' where appropriate.
We postulate however a generalized {\em discrete} \LR symmetry, 
which requires a representation content which is symmetric under 
the simultaneous exchange of ``L''~$\leftrightarrow$~``R'' and the
chiralities ``l''~$\leftrightarrow$~``r''. This generalization of
discrete \LR symmetry becomes necessary when the connection between 
the gauge groups and chiralities is not present. Discrete \LR symmetry
enforces as usual that the $SU(2)$ gauge couplings are identical:
$g_{2L} \equiv g_{2R} \equiv g_2$.

The so--called minimal \LR symmetric model contains a remarkable simple
set of possible irreducible representations. The gauge bosons are
$G^c_\mu$, $W^i_{L\mu}$, $W^i_{R\mu}$, $B_{\mu}$ for $SU(3)_c$,
$SU(2)_L$, $SU(2)_R$,  $U(1)_{B-L}$, respectively. The three known
generations of quarks and leptons fit very economically into the
discrete \LR symmetric set of representations $Q_L \sim (3,2,1,1/3)_l$,
$Q_R \sim (3,1,2,1/3)_r$, $\Psi_L \sim (1,2,1,-1)_l$ and $\Psi_R \sim
(1,1,2,-1)_r$, or more explicitly:
\beq
\label{lrfermionen}
\begin{array}{ccc}
Q_{L/R}= {u_c \choose d_c}_{l/r}~, 
       & {c_c \choose s_c}_{l/r}~, 
       & {t_c \choose b_c}_{l/r}~, \\
\\
\Psi_{L/R} =
             {\nu_e \choose e^-}_{l/r}~, 
           & {\nu_{\mu} \choose\mu^-}_{l/r}~, 
           & {\nu_{\tau} \choose \tau^-}_{l/r}~.
\end{array}
\eeq
The Higgs sector contains finally the scalar bi--doublet
\beq
\Phi \cong \left( \begin{array}{cc}
                     \Phi_1^0 & \Phi_1^+ \\
                     \Phi_2^- & \Phi_2^0
                  \end{array} \right)~,
\eeq
with $\Phi\sim (1,2,2,0)$ and two scalar triplets 
$\Delta_L$, $\Delta_R \sim (1,3,1,2)\oplus (1,1,3,2)$. 
Any \LR model with a larger or modified particle content will be called
{\em non--minimal}. 

We will consider new exotic fermions which can give rise to chiral
anomalies. We consider therefore only anomaly free combinations. 
The individual triangle anomaly contributions are proportional to
\beq
\label{defanomkoeff}
A^{abc} := \frac{1}{2} Tr \left[  \{ T^a, T^b \} T^c \right] ~.
\eeq
Here $T^a$ are the generators of the fermion representations which act
at the vertices. For abelian groups (as the $U(1)$ is) $T^a$ is just
a multiple of the unity matrix. If $A^{abc}$ vanishes for all indices
$a$, $b$, and $c$ then all higher order anomalies are absent, too. If
the fermions are given in a chiral basis with left-- and right--handed
states, $A^{abc}$ can be written as \cite{GeorgiGl}
\beq
\label{glanomlr}
A^{abc} = 2 \cdot \left( A_l^{abc} - A_r^{abc} \right) ~,
\eeq
where
\beq
A_{l,r}^{abc} := \frac{1}{2} Tr \left[  \{ T_{l,r}^a, T_{l,r}^b \}
T_{l,r}^c
\right] ~.
\eeq
Here $T_l^a$ and $T_r^a$ are the generators of the representations 
of the left--handed and right--handed fermions, respectively.

Choi and Volkas \cite{Choi} studied chiral anomalies which arise from
exotic fermions in \LR symmetric models. They showed that all chiral 
anomalies except those of the type $[SU(2)_L]^2 [U(1)_{B-L}]$ and
$[SU(2)_R]^2 [U(1)_{B-L}]$ cancel if the following two conditions are 
fulfilled: 
\begin{itemize}
\item[i)] The number of left--handed fermionic representations 
$(N_c, N_L, N_R, B-L+X)_l$ is equal to the number of the corresponding
right--handed representations \newline
\hbox{$(N_c, N_R, N_L, B-L+X)_r$}.
\item[ii)] The $U(1)$ quantum numbers of both representations are
equal.
\end{itemize}
Note that these conditions are automatically fulfilled in our case 
since they are equivalent to the generalized discrete \LR symmetry.
Eventually surviving $[SU(2)_L]^2 [U(1)_{B-L}]$ and 
$[SU(2)_R]^2 [U(1)_{B-L}]$ anomalies can be calculated by a simple 
formula which reads in our notation
\bea
\label{formchoi1}
A_{L,l}^{abc} \left( (N_c, N_L, N_R, b)_l \right) & = & N_c \cdot c_2
(T_l^{(N_L-dim.)}) \cdot N_R \cdot b ~, \\
\label{formchoi2}
A_{L,r}^{abc} \left( (N_c, N_L, N_R, b)_r \right) & = & N_c \cdot c_2 
(T_r^{(N_L-dim.)}) \cdot N_R \cdot b ~.
\eea
Here $A_L^{abc}$ and $A_R^{abc}$ are the triangle anomaly coefficients 
of the type $[SU(2)_L]^2 [U(1)_{B-L}]$ and $[SU(2)_R]^2 [U(1)_{B-L}]$. 
For left--handed (right--handed) representations 
$(N_c, N_L, N_R, b)_l$ ( $(N_c, N_L, N_R, b)_r$ ) 
the anomaly coefficients are called  $A_{L,l}^{abc}$ and
$A_{R,l}^{abc}$ ($A_{L,r}^{abc}$ and $A_{R,r}^{abc}$).
$T_{l/r}^{(N_L-dim.)}$ is a $N_L$-dimensional representation of
$SU(2)_L$, and $c_2$ is its ``index'' which we will define and discuss
later on in the context of the $\beta$--functions. For a singlet we
have $c_2 (1) = 0$. The coefficients $A_{R,l}^{abc}$ and
$A_{R,r}^{abc}$ can be easily obtained via
\bea
A_{R,l}^{abc} \left( (N_c, N_L, N_R, b)_l \right) & = &
A_{L,l}^{abc} \left((N_c, N_R, N_L, b)_l \right) ~, \\
A_{R,r}^{abc} \left( (N_c, N_L, N_R, b)_r \right) & = &
A_{L,r}^{abc} \left((N_c, N_R, N_L, b)_r \right) ~.
\eea
Equation~(\ref{glanomlr}) can now be rewritten as\footnote{Apparently,
the calculation of the different coefficients is independent of the
chosen generators of each representation. Thus we will omit from now
on the indices {$abc$}.}
\bea
A_L & = & 2 \cdot \left( A_{L,l} - A_{L,r} \right) ~, \\
A_R & = & 2 \cdot \left( A_{R,l} - A_{R,r} \right) ~.
\eea
The contributions to $A_L$ and $A_R$ are calculated for a number of
representations and are listed later in Table~\ref{tabexferm} together 
with their contributions to the $\beta$--functions. This will allow us
to search for anomaly free combinations of exotic fermions. 

The unification of running gauge couplings in the presence of 
exotic particles has been studied in supersymmetric extensions 
of the \LR model. In the literature are, however, only a few attempts 
of such studies which extend the minimal \LR symmetric model by some 
``exotic'' particles like higher--dimensional (not fundamental) 
irreducible representations. A search for anomalies in \LR models
was performed by Choi and Volkas \cite{Choi}. We use some of 
their results and combine it with a search for unification solutions.

The evolution of the gauge couplings $g_i$, $i=1,2,3$ is 
(for a given model with known particle content) governed by the 
corresponding $\beta$--functions 
\beq
\tabl{g_i(t)}{t} = \beta_{g_i}(g_j(t)) ~,
\eeq
where the dimensionless scale parameter is as usually defined 
through $t := \ln\left( \frac{\mu}{\mu_0} \right)$. 
To one--loop order $\beta_{g_i}$ has the form
\beq
\beta_{g_i} = \frac{(g_i)^3}{16 \pi^2} \cdot b_i ~,
\eeq
where $b_i$ is constant over some (or the whole) energy range.
Since we consider only weak couplings we can restrict ourselves to
one--loop calculations. Particle thresholds are described in first
approximation by the $\theta$--function approach with the assumption
that the new particles we introduce are not too wide spread around the
common scale $M_R$ which denotes the effective mass scale of the \LR
symmetric model. For the one--loop beta functions we use the following
general form given by Jones \cite{Jones}:
\bea
\beta_{g} & = & \frac{(g)^3}{16 \pi^2} \cdot \left[ ~ -\frac{11}{3} c_1
+ \frac{2}{3} \sum_{R_f} c_2 (R_f) \cdot d_1 (R_f) \cdot \dots \cdot
d_n (R_f) \right. \nonumber \\
 & & \left. + \frac{1}{3} ~\sum_{R_s} c_2 (R_s) \cdot d_1 (R_s) \cdot
\dots \cdot d_n (R_s) ~ \right] ~. \label{betaeq1}
\eea
The meaning of the notation is as follows: For a given group factor
$G_0$ we can write the full gauge group $G$ as a direct product in the
form
\beq
G = G_0 \otimes X_1 \otimes \dots \otimes X_n~,
\eeq
where $g$ is the coupling which belongs to $G_0$.
$R_f$ is then any irreducible chiral representation of fermions and
$R_s$ is any irreducible scalar representation of $G_0$. $d_i (R)$ is
the dimension of the representation $R$ under the group factor $X_i$.
The constants $c_2 (R)$ are fixed by the choice of representations and
depend on their dimensions. They are defined through
\beq
c_2 (R) \cdot \delta^{ab} := Tr [R^a R^b]~,
\eeq
and are equivalent to a normalization of the generators of $R$. $c_2
(R)$ is called the ``index'' of the irreducible representation $R$ and
is often also denoted by $T(R)$.\footnote{Unfortunately, the literature
contains some confusion in the naming of group--theoretical constants: 
the so--called ``second order Casimir invariant'', which we would
define by
\beq
c_3 (R) \cdot \delta_{ij} := \sum_a \left( R^a R^a \right)_{ij} =
\frac{dim(R^{adj})}{dim(R)} \cdot c_2 (R) ~,
\eeq
where $R^{adj}$ means the adjoint representation of $G_0$, is sometimes
also denoted by $c_2$. We adopted here the convention used in the book
by Bailin and Love \cite{BailinLove}.}
Furthermore we have as usual $c_1 = c_2(R^{adj})$ ($=N$ for $SU(N)$).

With eq.~(\ref{betaeq1}) the calculation of $\beta$--functions is
reduced to the task of finding $c_2 (R)$ for the relevant
representations. In our convention we have $Tr [R^a R^b] = \frac{1}{2}
\delta^{ab}$ and $c_2$ can be determined for any representation with
the help of the so--called ``Dynkin--labels'' \cite{Slansky,McKay}.
Note that the values given in refs.~\cite{Slansky,McKay} differ by
normalization conventions, in the case of ref.~\cite{Slansky} a factor 
2 and in the case of ref.~\cite{McKay} a factor $2\cdot rank(G)$. In
Tables~{\ref{tabsu2}} and {\ref{tabsu3}} we list some values of $c_2$
which we will use.  

\begin{table}[ht]
\begin{minipage}[t]{7.2cm}
\renewcommand{\arraystretch}{1.3}
\begin{tabular}[t]{|ccc|}
\hline
Dynkin--Label & Representation & $c_2$ \\ \hline
(1) &  2 & $\frac{\T 1}{\T 2}$ \\
(2) &  3 & 2 \\
(3) &  4 & 5 \\
(4) &  5 & 10 \\
(5) &  6 & $\frac{\T 35}{\T 2}$ \\
(6) &  7 & 28 \\
(7) &  8 & 42 \\
(8) &  9 & 60 \\
(9) & 10 & $\frac{\T 165}{\T 2}$ \\
(10) & 11 & 110 \\ \hline
\end{tabular}
\caption[Table of the $c_2$ for representations of the
$SU(2)$]{\label{tabsu2} $c_2 (R)$ for some irreducible representations
of $SU(2)$}  
\end{minipage}
\hfill
\begin{minipage}[t]{7.2cm}
\renewcommand{\arraystretch}{1.3}
\begin{tabular}[t]{|ccc|}
\hline
Dynkin--Label & Representation & $c_2$ \\ \hline
(1,0) &  3 & $\frac{\T 1}{\T 2}$ \\
(2,0) &  6 & $\frac{\T 5}{\T 2}$ \\
(1,1) &  8 & 3 \\
(3,0) & 10 & $\frac{\T 15}{\T 2}$ \\
(2,1) & 15 & 10 \\
(4,0) & 15 & $\frac{\T 35}{\T 2}$ \\
(0,5) & 21 & 35 \\
(1,3) & 24 & 25 \\
(2,2) & 27 & 27 \\
(6,0) & 28 & 63 \\ \hline
\end{tabular}
\caption[Table of the $c_2$ for representations of the
$SU(3)$]{\label{tabsu3} $c_2 (R)$ for some irreducible representations
of $SU(3)$}  
\end{minipage}
\end{table}

With this information it is easy to confirm for example the
coefficients $b_i$ of the one--loop $\beta$--functions of the minimal
\LR symmetric model to be 
\beq
\label{betaminlr}
\bvec{b_1}{b_2}{b_3}_{\bf LR} = \bvec{0}{-\frac{22}{3}}{-11} + N_G
\cdot \bvec{\frac{4}{3}}{\frac{4}{3}}{\frac{4}{3}} + N_{HBiD} \cdot
\bvec{0}{\frac{1}{3}}{0} + N_{HTr} \cdot \bvec{3}{\frac{2}{3}}{0} ~,
\eeq
where $N_G=3$ is the number of (complete) fermion generations,
$N_{HBiD}=1$ is the number of Higgs bidoublets, and $N_{HTr}=1$ is the
number of {\em pairs} of Higgs triplets. These coefficients will be
modified if we introduce some new (exotic) particles. We list therefore
in Table~\ref{tabexferm} contributions of exotic fermion
representations to the $\beta$--functions (together with the chiral
anomaly contributions $A_L$ and $A_R$ discussed above). Furthermore we
list in Table~\ref{tabexskalar} a number of scalar representations and
their contributions to the $\beta$--functions.

\begin{table}[thb]
\renewcommand{\arraystretch}{1.0}
\begin{center}
\begin{tabular}[t]{|c|c|lcl|lcl|}
\hline
Irreducible representation  &Contribution& \multicolumn{6}{c|}{Anomaly
coefficient} \\ \cline{3-8}
$(3_c,2_L,2_R,B-L+X)_{l,r}$ & $(b_1, b_2, b_3)$ & & $A_L$&  & & $A_R$&
\\ \hline 
$(1,1,1,b)_{l} \oplus (1,1,1,b)_{r}$ & ($\frac{\T 1}{\T 2} b^2$, 0, 0)
& &0& & &0& \\ \hline
$(3,1,1,b)_{l} \oplus (3,1,1,b)_{r}$ & ($\frac{\T 3}{\T 2} b^2$, 0,
\tf{4}{3}) & &0& & &0& \\ \hline
$(6,1,1,b)_{l} \oplus (6,1,1,b)_{r}$ & ($3 b^2$, 0, \tf{20}{3}) & &0& &
&0& \\ \hline
$(8,1,1,b)_{l} \oplus (8,1,1,b)_{r}$ & ($4 b^2$, 0, 8) & &0& & &0&
\\ \hline
$(10,1,1,b)_{l} \oplus (10,1,1,b)_{r}$ & ($5 b^2$, 0, 20) & &0& & &0&
\\ \hline
$(1,2,1,-1)_l \oplus (1,1,2,-1)_r$ & (1, \tf{1}{3}, 0) & l: -1 & \vline
& r: 0 & l: 0 & \vline &  r: 1 \\ \hline
$(1,2,1,b)_l \oplus (1,1,2,b)_r$ & ($b^2$, \tf{1}{3}, 0) & l: b &
\vline & r: 0 & l: 0 & \vline &  r: -b \\ \hline
$(1,2,2,b)_{l} \oplus (1,2,2,b)_{r}$ & ($2 b^2$, \tf{2}{3}, 0) & &0& &
&0& \\ \hline
$(1,3,1,b)_l \oplus (1,1,3,b)_r$ & ($\frac{\T 3}{\T 2} b^2$, \tf{4}{3},
0) & l: 4b &\vline& r: 0 & l: 0 &\vline& r: -4b \\ \hline
$(1,3,2,b)_l \oplus (1,2,3,b)_r$ & ($3 b^2$, \tf{8}{3}, 0) & l: 8b &
\vline& r: -3b & l: 3 b &\vline& r: -8b \\ \hline
$(1,3,3,b)_{l} \oplus (1,3,3,b)_{r}$ & ($\frac{\T 9}{\T 2} b^2$, 4, 0)
& &0& & &0& \\ \hline
$(3,2,1,\frac{1}{3})_l \oplus (3,1,2,\frac{1}{3})_r$ & (\tf{1}{3}, 1,
\tf{4}{3}) & l: 1 &\vline& r: 0 & l: 0 &\vline& r: -1 \\ \hline
$(3,2,1,b)_l \oplus (3,1,2,b)_r$ & ($3 b^2$, 1, \tf{4}{3}) & l: 3b &
\vline& r: 0 & l: 0 &\vline& r: -3b \\ \hline
$(3,2,2,b)_{l} \oplus (3,2,2,b)_{r}$ & ($6 b^2$, 2, \tf{8}{3}) & &0& &
&0& \\ \hline
$(3,3,1,b)_l \oplus (3,1,3,b)_r$ & ($\frac{\T 9}{\T 2} b^2$, 4, 2) &
l: 12b &\vline& r: 0 & l: 0 &\vline& r: -12b \\ \hline
$(3,3,2,b)_l \oplus (3,2,3,b)_r$ & ($9 b^2$, 8, 4) & l: 24b &\vline& r:
-9b & l: 9b &\vline& r: -24b \\ \hline
$(3,3,3,b)_{l} \oplus (3,3,3,b)_{r}$ & ($\frac{\T 27}{\T 2} b^2$, 12,
6) & &0& & &0& \\ \hline
\end{tabular}
\end{center}
\caption{\label{tabexferm} {\em Contributions of fermion
representations to the $\beta$--functions and anomaly coefficients. The
assignments of chirality ``l'' and ``r'' are essentially arbitrary. The
symbol ``$\oplus $'' denotes a pair of representations respecting the
generalized discrete \LR symmetry. The $B-L+X$ charge $b$ can take
different values for different pairs of representations; i.e., in
general the $b$'s differ from line to line.}}
\end{table}

\begin{table}[tbh]
\renewcommand{\arraystretch}{1.0}
\begin{center}
\begin{tabular}[t]{|c|c|}
\hline
Irreducible representation  & Contribution to   \\
$(3_c,2_L,2_R,B-L+X)$        & $(b_1, b_2, b_3)$ \\ \hline
$(1,1,1,b)$                 & ($\frac{\T 1}{\T 8} b^2$, 0, 0) \\ \hline
$(1,2,1,1)\oplus (1,1,2,1)$ & (\tf{1}{2}, \tf{1}{6}, 0) \\ \hline
$(1,2,1,b)\oplus (1,1,2,b)$ & ($\frac{\T 1}{\T 2} b^2$, \tf{1}{6}, 0)
\\ \hline
$(1,2,2,0)$                 & (0, \tf{1}{3}, 0) \\ \hline
$(1,2,2,b)$                 & ($\frac{\T 1}{\T 2} b^2$, \tf{1}{3}, 0)
\\ \hline
$(1,3,1,2)\oplus (1,1,3,2)$ & (3, \tf{2}{3}, 0) \\ \hline
$(1,3,1,b)\oplus (1,1,3,b)$ & ($\frac{\T 3}{\T 4} b^2$, \tf{2}{3}, 0)
\\ \hline
$(1,3,2,b)\oplus (1,2,3,b)$ & ($\frac{\T 3}{\T 2} b^2$, \tf{4}{3}, 0)
\\ \hline
$(1,3,3,b)$                 & ($\frac{\T 9}{\T 8} b^2$, 2, 0) \\ \hline
$(3,1,1,b)$                 & ($\frac{\T 3}{\T 8} b^2$, 0, \tf{1}{6})
\\ \hline
$(6,1,1,b)$                 & ($\frac{\T 3}{\T 4} b^2$, 0, \tf{5}{6})
\\ \hline
$(8,1,1,b)$                 & ($b^2$, 0, 1) \\ \hline
$(3,2,1,b)\oplus (3,1,2,b)$ & ($\frac{\T 3}{\T 2} b^2$, \tf{1}{2},
\tf{2}{3})\\ \hline
$(3,2,2,b)$                 & ($\frac{\T 3}{\T 2} b^2$, 1, \tf{2}{3})
\\ \hline
\end{tabular}
\end{center}
\caption{\label{tabexskalar} {\em Contributions of some scalar
representations to the $\beta$--functions.}}
\end{table}

\begin{table}[thb]
\begin{tabular}[thb]{|ccc|ccc|c|c|c|c|}
\hline
$F_1$ & $F_2$ & $F_3$ & $H_1$ & $H_2$ & $H_3$ & $M_R$ [GeV] & $M_{GUT}$
[GeV] & $\alpha_{GUT}^{-1}$ & $\tau_P$ [yrs] \\ \hline
 0  &  0  &  1 (1) &  0  &  1  &  1  & $2 \cdot 10^3$ & $10^{16.6}$ &
33.1 & $7.9 \cdot 10^{37}$ \\
 0  &  0  &  1 (1) &  0  &  1  &  2  & $1 \cdot 10^3$ & $10^{14.7}$ &
29.7 & $1.6 \cdot 10^{30} *$ \\
 0  &  0  &  1 (1) &  0  &  2  &  1  & $1 \cdot 10^6$ & $10^{15.9}$ &
34.6 &  $1.3 \cdot 10^{35}$ \\
 0  &  0  &  1 (1) &  1  &  1  &  0  & $2 \cdot 10^4$ & $10^{18.3}$ &
36.9 & $6.2 \cdot 10^{44}$ \\
 0  &  0  &  1 (1) &  1  &  1  &  1  & $5 \cdot 10^3$ & $10^{16.1}$ &
32.7 & $7.7 \cdot 10^{35}$ \\
 0  &  0  &  1 (1) &  1  &  2  &  0  & $5 \cdot 10^6$ & $10^{17.2}$ &
37.3 & $2.5 \cdot 10^{41}$ \\
 0  &  0  &  1 (1) &  2  &  1  &  0  & $5 \cdot 10^4$ & $10^{17.7}$ &
36.2 & $2.4 \cdot 10^{42}$ \\
 0  &  0  &  2 (\tf{2}{3})&  0  &  1  &  1  & $4 \cdot 10^4$ &
$10^{19}$ & 24.1 & $1.7 \cdot 10^{47}$ \\
 0  &  1 (\tf{5}{3}) &  0  &  1  &  1  &  0  & $2 \cdot 10^8$ &
$10^{15.3}$ &  42.8 & $8.4 \cdot 10^{32} *$ \\
 0  &  1 (\tf{2}{3}) &  1 (1) &  0  &  1  &  1  & $4 \cdot 10^6$ &
$10^{15.2}$ & 34.2 & $2.1 \cdot 10^{32} *$ \\
 0  &  1 (\tf{4}{3}) &  1 (\tf{2}{3}) &  1  &  1  &  0  & $1 \cdot
10^8$ & $10^{16.4}$ & 37.4 & $1.6 \cdot 10^{37}$ \\
 1 (\tf{5}{3}) &  0  &  0  &  0  &  1  &  1  & $1 \cdot 10^3$ &
$10^{16.5}$ & 42.7 & $5.3 \cdot 10^{37}$ \\
 1 (\tf{5}{3}) &  0  &  0  &  0  &  1  &  2  & $1 \cdot 10^3$ &
$10^{14.5}$ & 38.1 & $4.2 \cdot 10^{29} *$ \\
 1 (\tf{5}{3}) &  0  &  0  &  0  &  2  &  1  & $4 \cdot 10^5$ &
$10^{15.9}$ & 41.6 & $1.1 \cdot 10^{35}$ \\
 1 (\tf{5}{3}) &  0  &  0  &  1  &  1  &  0  & $6 \cdot 10^3$ &
$10^{18.4}$ & 47.2 & $2.6 \cdot 10^{45}$ \\
 1 (\tf{5}{3}) &  0  &  0  &  1  &  1  &  1  & $3 \cdot 10^3$ &
$10^{16}$ & 41.6 & $5.0 \cdot 10^{35}$ \\
 1 (2) &  0  &  0  &  1  &  2  &  0  & $1 \cdot 10^3$ & $10^{17}$ &
43.9 & $5.6 \cdot 10^{39}$ \\
 1 (\tf{5}{3}) &  0  &  0  &  2  &  1  &  0  & $2 \cdot 10^4$ &
$10^{17.7}$ & 45.7 & $3.8 \cdot 10^{42}$ \\
 1 (\tf{2}{3}) &  0  &  1 (\tf{2}{3}) &  0  &  1  &  1  & $4 \cdot
10^4$ & $10^{19}$ & 34.7 & $3.5 \cdot 10^{47}$ \\
 1 (\tf{4}{3}) &  1 (\tf{4}{3}) &  0  &  0  &  1  &  1  & $5 \cdot
10^4$ & $10^{14.9}$ & 39.3 & $1.8 \cdot 10^{31} *$ \\
 1 (\tf{4}{3}) &  1 (\tf{4}{3}) &  0  &  1  &  1  &  0  & $2 \cdot
10^5$ & $10^{16.2}$ & 42.4 & $3.3 \cdot 10^{36}$ \\
 2 (\tf{1}{3}) &  0  &  0  &  0  &  1  &  1  & $4 \cdot 10^6$ &
$10^{18.7}$ & 45.6 & $3.8 \cdot 10^{46}$ \\ \hline
\end{tabular}
\caption{\label{solutionstab} Unification solutions with less or equal 
4 extra representations. $F_1$, $F_2$, $F_3$ and $H_1$, $H_2$, $H_3$
are the numbers of fermions and scalars mentioned in the text. $F_4$
and $F_5$ are here always zero. The numbers in brackets in the
fermionic columns are the $U(1)$ charges. The table lists the LR--scale
$M_R$, the GUT scale $M_{GUT}$, the inverse coupling at $M_{GUT}$ and
the proton life time. A `$*$' after $\tau_P$ marks cases which are
close to the proton decay limit. } 
\end{table}

We have now all necessary ingredients to search for anomaly free 
combinations of exotic particles which lead to \LR gauge unification.
We define $\alpha_i=g_i^2/4\pi$ and evolve $\alpha_i^{-1}$ which
becomes simply
\beq
\frac{d\alpha_i^{-1}}{dt} = -~\frac{b_i}{2\pi}~.
\label{runinverse}
\eeq
In a first step the \SM is evolved with its known particle 
content to a common \LR scale $M_R$ where the \LR evolution starts. 
Eq.~(\ref{betaeq1}) allows us to determine the well known 
$\beta$--function coefficients for the \SM (with $N_G=3$ and with 
the $U(1)_Y$ coupling in GUT--normalization, i.e. multiplied by a 
factor $3/5$) 
\beq
\label{betaSM}
\bvec{b_{1Y}}{b_2}{b_3}_{\bf SM} = 
\bvec{+\frac{41}{10}}{-\frac{19}{6}}{-7}~.
\eeq
The experimental initial values for the gauge 
couplings at $M_Z$ are taken to be\footnote{The initial
$\alpha_{1,2}^{-1}$ are calculated in the usual manner from the LEP
data \cite{Hollik} for $\alpha_{QED}^{-1}(M_Z)$ and
$sin^2(\theta_W)(M_Z)$, where we have enlarged the $\alpha_{QED}^{-1}$
error bar by a factor of 2 due to the theoretical uncertainties. For
$\alpha_3(M_Z)$ we use the ``world average'' \cite{Bethke}
$\alpha_3(M_Z)_{world} = 0.117 \pm 0.006$.}:
\beq
\label{initial}
\bvec{\alpha_{1Y}^{-1}(M_Z)}{\alpha_2^{-1}(M_Z)}{\alpha_3^{-1}(M_Z)} = 
\bvec{59.41\pm0.08}{29.88\pm0.06}{\ 8.55\pm0.51}~.
\eeq
Beyond $M_R$ we continue to run the gauge couplings in a next step with
the \LR model for an arbitrary (but anomaly free) particle content to
the GUT--scale $M_{GUT}$. However, the different normalizations of the
$U(1)$ generators must be taken into account:
$\alpha_1^{-1} \equiv \alpha_{1,B-L}^{-1} = \frac{5}{2}\alpha_{1Y}^{-1}
- \frac{3}{2} \alpha_2^{-1}$. For the evolution of the \LR model we
need the coefficients $b_i$ which receive contributions from all
particles. First there are the contributions from all the gauge fields
which are indisputable. Next we always add the contributions of the
known fermions. Since the gauge fields are already included it is
obvious that any further contribution to $b_i$ must be positive. Thus
we obtain lower bounds for the coefficients $b_i$ if we just include
the gauge field and the known fermion contributions:
\beq
\label{bimin}
\bvec{b_{1,min}}{b_{2,min}}{b_{3,min}}_{\bf LR} = 
\bvec{4}{-\frac{10}{3}}{-7}~.
\eeq
There exist also upper bounds on $b_i$ in order to avoid Landau 
poles in the gauge couplings before the unification scale $M_{GUT}$ 
is reached. From the evolution equation~(\ref{runinverse}) follows
that for large enough $b_i$ the inverse 
coupling $\alpha_i^{-1}$ will run to zero before the GUT--scale is 
reached. In order to avoid such Landau singularities the coefficients 
$b_i$ must be limited from above:
\beq
\label{biLandau}
b_i \leq b_{i,max} =
\frac{2\pi\alpha_i^{-1}(M_R)}{\ln(M_{GUT}/M_R)}~.
\eeq
With $M_R= 1$~TeV, $10^{19}$~GeV~$ = M_{Planck} \ge M_{GUT} \ge
10^{15}$~GeV and the initial values eq.~(\ref{initial}) one finds for
example
\beq
\label{bimax}
\bvec{b_{1,max}}{b_{2,max}}{b_{3,max}}_{\bf LR} 
\ltap \bvec{20.55}{6.89}{2.55}~.
\eeq
There exist also severe restrictions on the differences 
$b_3-b_2$ and $b_3-b_1$ which come from the requirement
of unification of $\alpha_3$ with $\alpha_2$ and
$\alpha_3$ with $\alpha_1$ in an acceptable scale range. 
First, for phenomenological reasons, we must require 
$1$~TeV~$\le M_R \le M_{GUT}$. The GUT--scale itself should
lie below the Planck--scale $M_{Planck}=10^{19}$~GeV.
An additional strong lower bound for potential GUT--scales comes from
the proton lifetime. We do not discuss here any specific GUT model, but
simply demand that the GUT--scale is high enough to avoid such
problems. The proton life time is parameterized in terms of the usual
relation
\beq
\label{eqtaup}
\tau_{proton} \approx \frac{1}{(\alpha_{GUT})^2} \cdot
\frac{M_{GUT}^4}{M_{proton}^5}~,
\eeq
derived from an effective low energy four--fermion approximation for 
the decay channel $p \rightarrow e^+ + \pi^0$. For all potential GUT
scenarios with exotic particles we require that the experimental lower 
bound \cite{PDG} $\tau_P/B(p\rightarrow e^+ + \pi^0) > 9 \cdot
10^{32}~y$ is not badly violated. To fulfill this we demand $M_{GUT}
\ge 10^{15}$~GeV in our search. Solutions which lead to $M_{GUT}$ close
to this lower bound should be carefully checked if they lead to
problems with proton decay. In principle such a check should include
the discussion of decay modes, threshold corrections and higher orders
in the $\beta$--functions. From the above requirements on the involved
scales one finds severe restrictions on $b_3-b_2$ and $b_3-b_1$. 
For $M_R = 1$~TeV one finds for example $-4.33 \le b_3-b_2 \le -3.25$ 
and $ -20 \le b_3-b_1 \le -15$.

The above constraints on the coefficients $b_i$ lead to restrictions on
the possible representation content of GUT solutions. Since larger
representations contribute via eq.~(\ref{betaeq1}) systematically more
to the beta functions we can derive the largest representation which
does not lead to a Landau pole. This largest representation is given by
the requirement that the additional contributions to $b_i$ are smaller
than $b_{i,max}-b_{i,min}$. So we must exclude from
Table~{\ref{tabexferm}} the entries $(10,1,1,b)_{l} \oplus
(10,1,1,b)_{r}$ and $(3,3,3,b)_{l} \oplus (3,3,3,b)_{r}$ since the
limits for $b_2$ and $b_3$, respectively, are exceeded. We further
exclude $(6,1,1,b)_{l} \oplus (6,1,1,b)_{r}$, $(8,1,1,b)_{l} \oplus
(8,1,1,b)_{r}$, and $(3,3,2,b)_{l} \oplus (3,2,3,b)_{r}$, even though
the limits in $b_2$ and $b_3$ are not yet reached. It would however be
impossible to add any other fermionic representations which would
respect these limits and the one of $b_3 - b_2$ and lead to
unification. Some of the remaining representations should occur only in
``natural'', anomaly free combinations, namely $(1,2,1,b_a)_l \oplus
(1,1,2,b_a)_r$ together with $(3,2,1,b_b)_l \oplus (3,1,2,b_b)_r$,
which is a slight variation of one standard fermion generation, and
$(1,3,1,b_c)_l \oplus (1,1,3,b_c)_r$ together with $(3,3,1,b_d)_l
\oplus (3,1,3,b_d)_r$. Another ``natural'' combination would be
$(1,3,2,b_e)_l \oplus (1,2,3,b_e)_r$ together with $(3,3,2,b_f)_l
\oplus (3,2,3,b_f)_r$, but since we have already excluded the latter we
must also exclude $(1,3,2,b_e)_l \oplus (1,2,3,b_e)_r$. The signs of
the $U(1)$--charges $b$ must be opposite in each of the remaining
anomaly free pairs. Anomaly cancellation can be achieved by raising the
number of the ``smaller'' partner or by adjusting the $b$'s adequately
or both. We should also mention that the pair of singlets
$(1,1,1,b)_{l} \oplus (1,1,1,b)_{r}$ could be used for purposes of
``fine--tuning'' the running of the $U(1)$--coupling because it only
affects $b_1$ via $b$. We will however not take this possibility into
account.

There exist combinations of representations which do not affect 
$b_3-b_2$ and $b_3-b_1$. The slope of all $\alpha_i^{-1}$ is then 
changed by the same amount which lowers the value of
$\alpha_{GUT}^{-1}$, but does not affect unification. Thus if one has
one unification 
solution one can easily obtain another by adding such a combination 
as long as none of the other bounds is violated. Since one of the 
$b_i^\prime$s will eventually violate the bounds eq.~(\ref{biLandau})
due to Landau poles one 
obtains in this way only a finite number of extra solutions. A well
known combination of representations of the type mentioned above is a
complete extra generation 
of standard quarks and standard leptons which can be written as 
$\left( (1,2,1,-1)_{l} \oplus (1,2,1,-1)_{r} \right )\oplus
\left( (3,2,1,\frac{1}{3})_{l}\oplus (3,2,1,\frac{1}{3})_{r}\right )$.
For our case at most nine generations are allowed if one starts with
gauge fields and known fermions.
If additional fields (like the scalars of the minimal model) 
are present this number will be even smaller.

We can now perform a systematic search for GUT solutions. 
We use only those fermionic and scalar representations of 
Tables~{\ref{tabexferm}} and \ref{tabexskalar} 
which are still viable and which do affect unification.
We include the following fermions: $F_1$ copies of 
$(3,1,1,b)_{l} \oplus (3,1,1,b)_{r}$, $F_2$ copies of 
$(1,2,2,b)_{l} \oplus (1,2,2,b)_{r}$, $F_3$ copies of  
$(3,2,2,b)_{l} \oplus (3,2,2,b)_{r}$, $F_4$ (anomaly free) copies of 
$(1,3,1,b)_l \oplus (1,1,3,b)_r$ together with 
$(3,3,1,b)_l \oplus (3,1,3,b)_r$, and $F_5$ copies
of $(1,3,3,b)_{l} \oplus (1,3,3,b)_{r}$, where the $b$'s can be chosen
as required. Furthermore we include $H_1$ Higgs doublets $(1,2,1,1)
\oplus (1,1,2,1)$, $H_2$ Higgs bidoublets $(1,2,2,0)$, and $H_3$ Higgs 
triplets $(1,3,1,2)\oplus (1,1,3,2)$. For these scalars we fixed the
$U(1)$ charges to the familiar values in order to avoid ``too exotic''
particles. 

\clearpage 

Dropping this constraint would of course lead to many more solutions.
We demand, however, that we have at least one scalar bidoublet and at
least one pair of scalar doublets or scalar triplets in order that the
Higgs content allows a phenomenological viable \LR breaking sequence
$SU(2)_L\otimes SU(2)_R\otimes U(1)_{B-L} \rightarrow SU(2)_L \otimes
U(1)_Y \rightarrow U(1)_{QED}$\footnote{Note that this criterion
applies even to models like in \cite{ALS} where the Higgs sector is
composite and where the scalars contribute only effectively to the
running of the gauge couplings.}. We
present in Table~\ref{solutionstab} the ``most minimal solutions''
of our search where the total number of additional fermionic and scalar
representations is required to be less than or equal to 4. $F_4$ and 
$F_5$ are not listed since they are always zero if less than 5 extra
representations are required.
The first six columns of Table~\ref{solutionstab} contain the numbers
of fermionic ($F_1$, $F_2$, $F_3$) and scalar ($H_1$, $H_2$, $H_3$)
representations, respectively. For the fermions we also list in
brackets the required $U(1)$ charge $b$. The next two columns list
the required \LR scale $M_R$ and the resulting GUT--scale. For
completeness, we give the values of $\alpha_{GUT}^{-1}$ and the proton
life time $\tau_P$ in the last two columns. As can be seen there are
solutions for a wide range of $M_R$, beginning with $10^3$~GeV up to
$10^8$~GeV. Also the GUT--scale varies from $\approx 10^{15}$~GeV to
$10^{19}$~GeV.

Apparently, there are many more solutions if we skip the condition that
at most 4 additional representations are used. In an extended search we
removed the constraint of less or equal than 4 new representations, but
we kept the set of scalars with fixed $U(1)$ charges only. However, we
omit all solutions where only extra scalar and no fermionic
representations are involved. Though these are GUT solutions, too, they
appear not interesting to us because they do not lower the scale $M_R$.
Some even raise it up to $M_R \approx 10^{11}$~GeV. Even though the
number of solutions of this search is bounded from above due to the
upper bounds on $b_i$ there are still many solutions. An unconstrained
computer search resulted in more than 1500 candidate solutions. 

\begin{figure}[th]
\rotate[r]{
\epsfysize=15cm
\epsffile{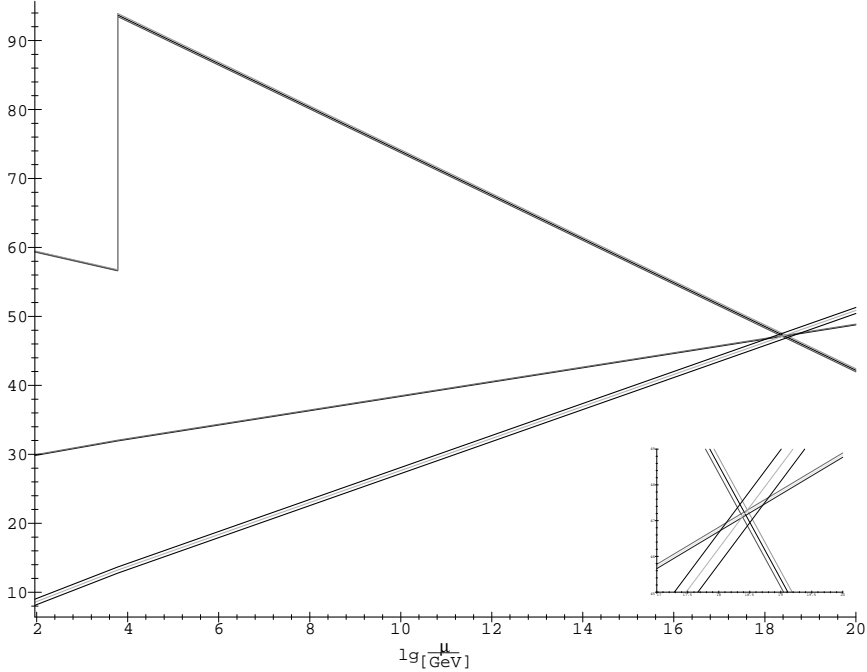}}
\caption{\protect{\label{solplot}} \em Running couplings in the
extended \LR symmetric model with one extra pair of exotic fermion
representations \hbox{$(3,1,1,\frac{5}{3})_{l} \oplus
(3,1,1,\frac{5}{3})_{r}$} and the Higgs sector of the doublet model,
i.e. one scalar bidoublet and one pair of doublets. The \LR scale is
\hbox{$M_R = 6 \cdot 10^3$}~GeV, and the GUT--scale \hbox{$M_{GUT} =
10^{18.4}$}~GeV.}
\end{figure}

In Fig.~\ref{solplot} we show as an example the solution $F_1=1$,
$H_1=1$, $H_2=1$ explicitly since it has very interesting properties.
In this example we have thus one scalar bidoublet, one pair of doublets
(the doublet model), and one additional fermionic pair of
$(3,1,1,\frac{5}{3})_{l} \oplus (3,1,1,\frac{5}{3})_{r}$. This solution
has a remarkable low \LR scale of only 6~TeV and a very high GUT--scale
which is almost the Planck scale. The proton life time $\tau_P \approx
10^{45}$~yrs is far beyond the lower experimental bound and thus
completely safe. One may ask whether such solutions with a very high
$M_{GUT}$ are physical meaningful or not. Taken seriously they could
point to a scenario which is usually not considered: New physics beyond
$M_R$ at the Planck scale could make a direct transition from Planck
scale physics to the \LR model {\em without} any intermediate stage.
The hierarchy problem might be solved when the Planck--scale physics is
expressed explicitly, but this information might also be lost
completely in the low energy effective Lagrangian.

Our search was performed at the one--loop level and thresholds were
described in the $\theta$--function approach with one common threshold
scale. We estimate that the corrections from higher orders and 
threshold functions are of the order of $1\sigma$. The scale $M_R$ was 
determined always as the ``best fit'', i.e., intersection of the
central lines. Since $M_R$ is treated as a free parameter it seems
always possible to absorb corrections into the precise value of $M_R$
as long as it is not too close to its lower or upper bound. However,
the analysis of unification solutions becomes much more involved if the
mass spectrum of the \LR model were widely spread around $M_R$. In this
case special attention has to be paid for example to those solutions
where $\tau_P$ is near the lower bound. Two--loop and threshold
corrections can change the result for $\tau_P$ up to one order of
magnitude.

In summary we have found many new possibilities of
non--supersymmetric \LR models with gauge coupling unification.
We required that the models are anomaly free and that the GUT--scale is
below the Planck--scale, but high enough to avoid problems with proton
decay. The \LR scale should lie between 1~TeV and the GUT--scale. 
Solutions were found even with very low \LR scale which was 
so far unknown. We presented a number of ``minimal''
solutions where the number of extra fermion and scalar representations
beyond the known fermions (quarks and leptons) and gauge fields was 
restricted to four. There exist strong upper limits on the size of 
representations in order to avoid Landau singularities before the
GUT--scale is reached. We demonstrated how this can be used to
eliminate 
large representations from the search. But without limitations on the
number of extra representations there are still many GUT--solutions.
In an unconstrained search we found more than 1500 candidates.
We think that this demonstrates that coupling unification alone 
is not very helpful to decide among candidates of new physics 
as long as one does not have a criterion for the type and number
of representations.

Some of the listed models with a rather small set of
``exotic particles''
may be worth to be studied phenomenologically in more detail. 
In these studies additional investigations on the behaviour of 
Higgs-- and Yukawa--couplings should be included.
It would also be interesting to include systematically one 
or two more intermediate stages on the way to the GUT--scale. 
One could for example consider extensions of the \LR model 
which contain an $SU(4) \otimes SU(2) \otimes SU(2)$, originally
proposed by Pati and Salam \cite{PatiSalam}.
Furthermore, the impact of higher dimensional operators suppressed by
the Planck scale which arise from quantum gravity or 
Kaluza--Klein--type theories can be taken into account. It is
interesting to see that this is still compatible with a low $M_R$
\cite{Datta}.

We are grateful to E.~Akhmedov and E.~Schnapka for comments and
useful discussions. 
This work was supported in part by the DFG grant Li519/2--1
and EC grant ERB~SC1*CT000729.

\end{document}